# High reflection emission assisted by phase compensation via anomalous dispersion

Hai-Tao Zhou[1], Miao-jun Guo[1], Da-Wei Wang[2], Jun-Xiang Zhang[1*], Shi-Yao Zhu[1,2,3]

[1] The State Key Laboratory of Quantum Optics and Quantum Optics Devices, Institute of Opto-Electronics, Shanxi University, Taiyuan 030006, China
[2] Beijing Computational Science Research Center, Beijing 100084, China
[3] Department of Physics, Hong Kong Baptist University, Hong Kong, China
*Corresponding author: junxiang@sxu.edu.cn

*The reflection spectrum of a probe light in a $\Lambda$-type three-level atomic system coupled by an off-resonant standing-wave is investigated experimentally and theoretically. We show that the maximum value of reflection coefficient occurs when both of the coupling and probe lights are tuned off resonances from the atomic transitions. The nature of enhanced reflection is attributed to the phase compensation caused by the anomalous dispersion, which leads to a significant reduction of nonlinear phase mismatch in atomic four-wave mixing. At certain detuning of coupling and probe frequencies near two-photon resonance, there exits a best compensation, so the reflection efficiency reaches its maximum. The dependences of efficiency on the intensity of coupling fields and the density of atoms are also studied.*

Recently there has been a great interest in the manipulation of dispersion [1,2] and absorption properties [3] of an absorbing medium for changing light propagation from subluminal to superluminal [4]. In a simple $\Lambda$-type three-level system, the absorption and dispersion of the probe light in a medium will be modified from the electromagnetically induced transparency (EIT) and normal dispersion [5] to the electromagnetically induced absorption (EIA) and anomalous dispersion [6,7], when the co-propagating coupling beam is replaced by a coupling standing-wave. EIT and EIA are two opposite atomic coherence effects, the EIT can result in high efficient nonlinear effects due to the transparency [8,9], and EIA may prohibit atoms from generation of nonlinearity due to the high absorption [10,11], however, the similar nonlinear multiwave mixing in EIA system can still be observed, and it was interpreted as electromagnetically induced absorption grating (EIG) generated by the standing wave of the coupling field [12-14]. In the Bragg reflection/diffraction direction, there is no original field, so that the detection sensitivity can be high [15], which may have potential application in quantum information processing and quantum network proposed in atomic ensembles [13,16].

The EIG resulted from a standing wave has been experimentally demonstrated in cold atoms and atomic vapor cells [15-18]. It has also been extensively studied through four-wave mixing (FWM) in atoms [9,19,20], where the generated FWM signal can also be considered as the Bragg scattering. A recently study on stationary light pulses (SLPs) based on the atomic coherence showed that the Bragg grating was not the necessary condition for such kind of multi-wave processes [21], where the mechanism of the multi-wave mixing was not the EIG. Most of the EIG or EIG-related FWM experiments focused on the atomic scheme driven by a resonant standing wave to the atomic transition, since it was examined that the EIG efficiency would be reduced with a off-resonant standing wave [17], however there was an unexplained reduction or increase of efficiency in a system coupled with different detuned standing wave [16]. Here we present an experimental investigation of the off-resonant reflection of probe light in three-level atomic vapor with a detuned coupling standing wave, the experimental and simulation results are in a good agreement and indicate for the first time, that the obtained high efficiency of off-resonant reflection is due to the phase compensation caused by steep anomalous dispersion in coherently prepared EIA medium.

The experiment is performed in a $^{133}$Cs vapor cell of length L=7.5cm with AR coated end windows (the loss of the far-off resonant light through the cell is measured to be 4%). The D1 line is used to form a $\Lambda$-type three-level system, see Fig.1a, the two ground states are labeled as $|b\rangle$ ($6^2S_{1/2}, F=4$) and $|c\rangle$ ($6^2S_{1/2}, F=3$), and the excited state is $|a\rangle$ ($6^2P_{1/2}, F'=4$), The transition frequencies from $|a\rangle$ to $|b\rangle$ and $|c\rangle$ are $\omega_{ab}$ and $\omega_{ac}$, respectively, with $\Delta\omega = \omega_{ac} - \omega_{ab} = 9.2GHz$. The decay of the upper state to ground states is $\Gamma_a = 2\pi \times 4.6MHz$. The strong standing wave $\omega_c$ (parallel polarization) drives the transition between $|a\rangle$ and $|c\rangle$ with the detuning $\Delta_c = \omega_c - \omega_{ac}$, and the weak probe light $\omega_p$ (vertical polarization) couples the transition between $|a\rangle$ and $|b\rangle$ with the detuning $\Delta_p = \omega_p - \omega_{ab}$. Here our design makes the frequency of the coupling field is larger than the probe frequency (see Fig.1a), so that the Bragg condition, $\omega_c/\omega_p = \cos\varphi$ ($\varphi$ real), can't be fulfilled (Fig. 1b). However, high efficiency (up to 50%) for the signal is observed in the reflection direction (green dashed line in Fig.1c) of the probe. The mechanism is found in the following that the anomalous dispersion of the EIA system contributes compensation to the phase mismatching in a certain degree.

Fig.1c shows the experimental scheme, the standing wave is created by two counter-propagating coupling beams (formed by a high reflection mirror M), each of the coupling beams has a power of 20mW, and the co-propagating probe light has the power of 90μW. The e$^{-2}$ full width of probe and coupling beams are the order of 0.59mm and 0.64mm respectively. The co-propagating coupling and probe beams are incident in the cell forming an angle of $\theta = 0.14°$.

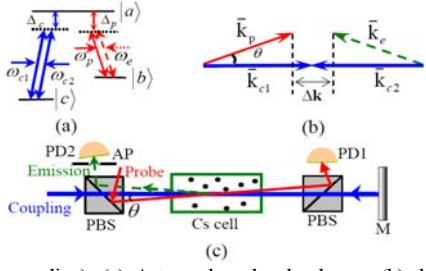

Fig. 1. (Color online): (a) Λ-type three-level scheme; (b) the sketch of phase matching; (c) Experimental setup. PBS, polarizing beam splitter; M, reflection-mirror; PD1,2, photo detectors; AP, aperture.

If we remove reflection mirror M in Fig.1c, a typical two-photon resonance EIT in Λ-type three-level system is observed, and a steep normal dispersion is accompanied. In this case, there is no signal be observed by PD2. However, when M is turned on, the EIT is transformed into the EIA and the dispersion of the system changes from normal to anomalous, meanwhile a reflection signal of the probe is observed by PD2. In Fig.2, the normalized reflection efficiency $\eta = P_e / P_p \times 100\%$ (emission power $P_e$ over probe power $P_p$) versus probe detuning is plotted for different detuning of coupling field. Note that the reflection peak occurs at $\Delta_p = -4.7 MHz$ even when the coupling frequency is on-resonance ($\Delta_c = 0$), see the black curve in Fig.2. When the coupling field is red detuned ($\Delta_c < 0$), it is seen that the height of the efficiency peak and its frequency detuning increase. The efficiency reaches the maximum at $\Delta_c = -11 MHz$. When the coupling is further red-detuned, the efficiency peak decreases. On the contrary, if the coupling field is blue-detuned ($\Delta_c > 0$), the efficiency peak decreases rapidly with $\Delta_c$ increase, see green and red lines in Fig.2.

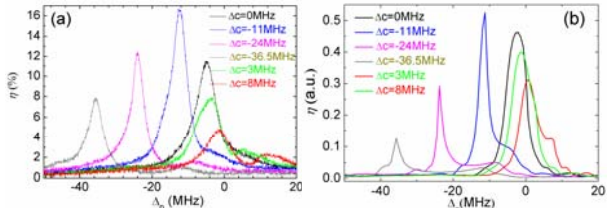

Fig.2. (Color online) emission spectrum of signal vs probe detuning with different detuning of coupling fields: (a) for experimental results with the temperature of vapor cell is T=43°C; (b) for theoretical curves with the parameters $\gamma_{ab} = 0.5\Gamma_a$, $\gamma_{cb} = 0.03\Gamma_a$, $\Omega_p = 2\pi \times 3 MHz$, $\Omega_{c1} = \Omega_{c2} = 2\pi \times 50 MHz$, the density of atoms $10^{15}/m^3$.

Motivated by experimental results, we consider a theoretical model as shown in Fig.1a, the equations of motion for the density matrix elements are

$$\dot{\rho}_{ab} = (-i\omega_{ab} - \gamma_{ab})\rho_{ab} + \frac{i}{2}(\Omega_{c1}e^{-i\omega_{c1}t} + \Omega_{c2}e^{-i\omega_{c2}t})\rho_{cb} + \frac{i}{2}\Omega_p e^{-i\omega_p t} \quad (1a)$$

$$\dot{\rho}_{cb} = (-i\omega_{cb} - \gamma_{cb})\rho_{cb} + \frac{i}{2}(\Omega_{c1}^* e^{i\omega_{c1}t} + \Omega_{c2}^* e^{i\omega_{c2}t})\rho_{ab} \quad (1b)$$

where $\Omega_p = (\mu_{ab}E_p/\hbar)e^{i\vec{k}_p \cdot \vec{r}}$, $\Omega_{c1,c2} = (\mu_{ab}E_{c1,c2}/\hbar)e^{i\vec{k}_{c1,c2} \cdot \vec{r}}$ are the Rabi frequency of probe field, co- and counter-propagating coupling fields, and $\mu_{ab}, \mu_{ac}$ are the relevant dipole moments, $E_p, E_{c1,c2}$ are the amplitude of the fields. $\gamma_{ab}, \gamma_{cb}$ are decay rates. Note that the wavevectors of the two coupling field are in opposite directions, $\vec{k}_{c1} = -\vec{k}_{c2} = \vec{k}_c$, and the frequencies have $\omega_{c1} = \omega_{c2} = \omega_c$. In Eq.(1), we have made the approximations that $\rho_{ab} = \rho_{cc} \cong 0, \rho_{bb} = 1$ and $\Omega_p \ll \Omega_{c1,c2}$. The solution $\rho_{ab}(t)$ (to first order of $E_p$ and all order of $E_{c1,c2}$) can be found in the form,

$$\rho_{ab}(t) = \sum_n \tilde{\rho}_{ab}^{[n]} e^{-i\omega_p t} e^{i(\vec{k}_p \cdot \vec{r} - 2n\vec{k}_c \cdot \vec{r})}, \quad (n = 0, \pm 1,...) \quad (2)$$

Here $\tilde{\rho}_{ab}^{[n]}$ represents the nth order emission induced by the medium. The zero-th order $\tilde{\rho}_{ab}^{[0]}$ corresponds to the linear absorption and dispersion of the probe field. The emission frequency of $\tilde{\rho}_{ab}^{[n]}$ is $\omega_e^{[n]} = \omega_p - (\vec{k}_p - \vec{k}_e - 2n\vec{k}_c) \cdot \vec{v}$, which depends on the atomic velocity, so that the n-th order emission ($n \neq 0$) from different velocity atoms has different frequency, the contribution of $\tilde{\rho}_{ab}^{[n]}$ (with $n \neq 0,1$) is small due to the dependence of frequency on the atom velocity. For n=1, the emission frequency is $\omega_e = \omega_e^{[1]} = \omega_p - (\vec{k}_p - \vec{k}_e - 2\vec{k}_c) \cdot \vec{v}$, which is equal to $\omega_p$ in the reflection direction when the phase matching condition $\Delta\vec{k} = \vec{k}_p - \vec{k}_e - 2\vec{k}_c = 0$ is satisfied, The reflection field emitted by atoms with the wave vector $\vec{k}_e$ and frequency $\omega_e$ is

$$E(\vec{k}_e, t) \propto \tilde{\rho}_{ab}^{[1]} \exp[\Delta\vec{k} \cdot \vec{r}] \quad (3)$$

The involved process can be described by the transitions $|b\rangle \rightarrow |a\rangle \rightarrow |c\rangle \rightarrow |a\rangle \rightarrow |b\rangle$ companied by absorbing the probe photon first, emitting the forward coupling photon second, then absorbing the backward coupling field photon, and finally emitting the photon of the reflected field, which are determined by energy and momentum conservation: $\omega_{c1} - \omega_{c2} = \omega_p - \omega_e$ and $\vec{k}_{c1} - \vec{k}_{c2} = \vec{k}_p - \vec{k}_e$ here $|\vec{k}_{c1}| = |\vec{k}_{c2}| = n_c \omega_c / c$, $|\vec{k}_p| = n_p \omega_p / c$ and $|\vec{k}_e| = n_e \omega_e / c$. In our experimental scheme, the energy conservation leads to the frequency relation $\omega_p = \omega_e$. We set $n_c = 1$ because the dispersion and absorption of strong coupling field can be neglected, and $n_p = n_e = 1 + (1/2)\text{Re}\,\tilde{\rho}_{ab}^{[0]}$ for both the probe and emission light are coupled to the same atomic transition. The value of $\Delta\vec{k}$ is ( see Fig. 1b)

$$\begin{aligned}|\Delta \vec{k}| &= 2(n_p \omega_p \cos\theta - \omega_c)/c \\ &= [2(\omega_p \cos\theta - \omega_c) + \text{Re}\,\tilde{\rho}_{ab}^{[0]}\omega_p \cos\theta]/c\end{aligned} \quad (4)$$

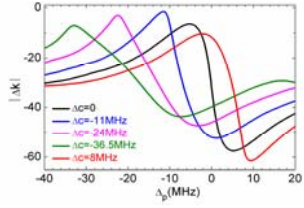

Fig. 3. Phase matching at $\theta = 0.14°$ vs the probe detuning for different detunings of coupling field. The parameters are the same as in Fig. 2.

Note that the first term on the right-hand side of Eq.(4) is $2(\omega_p \cos\theta - \omega_c)/c < 0$ because of $\omega_p < \omega_c$. The phase mathing condition will be satified ($|\Delta \vec{k}| \approx 0$) when the second term $[\text{Re}\,\tilde{\rho}_{ab}^{[0]}\omega_p \cos\theta]/c > 0$, which contributes phase compenstion by the dispersion $\text{Re}\,\tilde{\rho}_{ab}^{[0]}$. Fig.3 shows the theoretical predication for compensted $|\Delta \vec{k}|$, it is clearly seen that the compensation of $|\Delta \vec{k}|$, which lets $|\Delta \vec{k}|$ be increased to 0, can be significant when $\Delta_c < 0$. The optimum compensation exists when $\Delta_c \approx \Delta_p = -11 MHz$ (see blue line), for which $|\Delta \vec{k}|$ is almost close to 0. The existence of an optimum compensation can be understood from the dispersion propertie of the medium in Fig.4, which is measured using homodyne method based on Mach-Zehnder interferometer to detect the phase variation $\Delta\Phi$ of probe light passing through the cell.

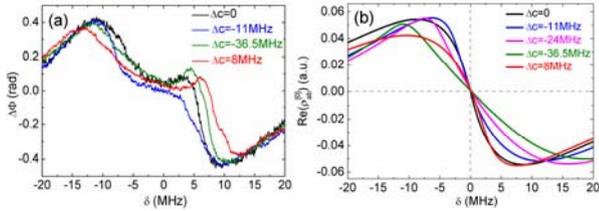

Fig. 4. (Color online) Phase shift of probe light vs two-photon detuning $\delta = \Delta_p - \Delta_c$ with different detunings of coupling field. The other experimental parameters are the same as in Fig. 2.

Because of driving by strong standing wave, the atomic medium exhibits a strong absorption and a steep anomalous dispersion for the probe and generated signal, and hence the dispersion for probe and reflected fields is anomalous. It is possible to reduce the phase mismatching by adjusting $\omega_p$, for smaller $\omega_p < \omega_{ab}$ ($\Delta_p < 0$) we have $n_p > 1$, see the curves in Fig.4. Furthermore, we can see that absolute value of the dispersion slope becomes larger when $\Delta_c$ shifts from 0 to $-11MHz$ (blue line in Fig.3), and it decreases for $\Delta_c < -11MHz$. That is to say, at $\Delta_c = -11MHz$, the reflection efficiency reaches its maximum, as shown in Fig. 2, 3. Contrarily, when $\Delta_c > 0$, the anomalous dispersion decreases with increase of $\Delta_c$ (red line in Fig.3), so the peak height of the reflection signal decreases monotonously, as shown in Fig.2.

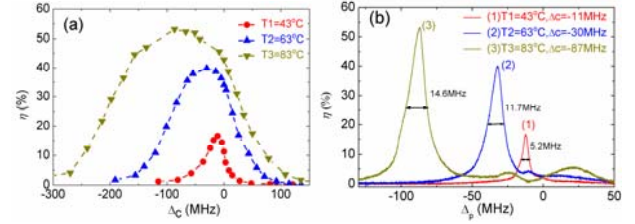

Fig. 5. (Color online) (a) the efficiency of reflection peaks vs the detuning of coupling field at three different temperatures: 43°C (red), 63°C (blue) and 83°C (dark yellow); (b) the emission spectrum with the three coupling detuning for the maximum efficiency.

In Fig.5, we measure the maximum efficiency versus $\Delta_c$ at three different temperatures, 43°C, 63°C and 83°C. Higher temperature corresponds to the higher atom density, which results in larger absorption and dispersion. Consequently, the maximum efficiency increases with the temperature. Note that the peak is shifted from -11MHz to -87MHz as the temperature increases. More than 20% efficiency can still be obtained even at large detuning of the coupling light (in the range of 100MHz), which might be useful to make the optical switch be tunable in a wide range controlled by the detuning of the coupling light. The efficiency peaks are at $\Delta_c$ = -11MHz, -30MHz and -87MHz for 43°C, 63°C, and 83°C, respectively, it is saturated above 85°C. And meanwhile, the FWHMs of spectrum are 5.2MHz, 11.7MHz and 14.5MHz at 43°C, 63°C, and 83°C, respectively (a clear broaden), which is caused primarily by the widen of collision line width.

In conclusion, we experimentally investigated the spectrum of the reflected field in three-level system coupled by strong standing-wave light. The experiments show that the highest efficiency is at near two photon resonance, but not at the single photon resonance. For the high efficiency observed here, the anomalous dispersion plays a key role for phase compensation. The reflected intensity reaches more than 50% of the input probe light at 83°C. Owing to the properties, for instance, high reflection efficiency, wide range of detuning of coupling light and without original field, the discussion may have many potential applications in quantum optics and quantum information, such as the multi-channels information processing, tunable optical switching.

Acknowledgements: This work is supported in part by the NSFC (No. 10974126, 60821004), National Basic Research Program of China (No. 2010CB923102), RGC of HK Government, and FRG of HKBU.